\PassOptionsToPackage{prologue,table}{xcolor}
\documentclass[acmtog]{acmart}

\usepackage{booktabs} %

\definecolor{best1}{RGB}{222,242,212}
\definecolor{best2}{RGB}{255,250,212}

\usepackage[ruled]{algorithm2e} %

\SetAlFnt{\small}
\SetAlCapFnt{\small}
\SetAlCapNameFnt{\small}
\SetAlCapHSkip{0pt}

\usepackage{multirow}
\usepackage{colortbl}
\usepackage[table]{xcolor}
\acmJournal{TOG}

\newcommand{\jyh}[1]{#1}

\setcopyright{rightsretained} 
\acmJournal{TOG}
\acmYear{2024} \acmVolume{43} \acmNumber{6} \acmArticle{} \acmMonth{12}\acmDOI{10.1145/3687926}

\begin{document}

\title{Robust Dual Gaussian Splatting for Immersive Human-centric Volumetric Videos}

\author{Yuheng Jiang}
\orcid{0000-0001-8121-0015}
\affiliation{%
	\institution{ShanghaiTech University}
	\city{Shanghai}
	\country{China}}
\affiliation{%
	\institution{NeuDim Digital Technology (Shanghai) Co.,Ltd.}
	\country{China}}
\email{zhaofq@shanghaitech.edu.cn}

\author{Zhehao Shen}
\affiliation{%
	\institution{ShanghaiTech University}
	\city{Shanghai}
	\country{China}
}
\email{shenzhh@shanghaitech.edu.cn}

\author{Yu Hong}
\affiliation{%
	\institution{ShanghaiTech University}
	\city{Shanghai}
	\country{China}
}
\email{hongyu@shanghaitech.edu.cn}

\author{Chengcheng Guo}
\affiliation{%
	\institution{ShanghaiTech University}
	\city{Shanghai}
	\country{China}
}
\email{guochch@shanghaitech.edu.cn}

\author{Yize Wu}
\affiliation{%
	\institution{ShanghaiTech University}
	\city{Shanghai}
	\country{China}
}
\email{wuyize25@163.com}

\author{Yingliang Zhang}
\affiliation{%
	\institution{DGene Digital Technology Co., Ltd.}
	\country{China}
}
\email{yingliang.zhang@dgene.com}

\author{Jingyi Yu}
\affiliation{%
	\institution{ShanghaiTech University}
	\city{Shanghai}
	\country{China}
}
\email{yujingyi@shanghaitech.edu.cn}

\author{Lan Xu}
\affiliation{%
	\institution{ShanghaiTech University}
	\city{Shanghai}
	\country{China}
}
\email{xulan1@shanghaitech.edu.cn}
\authornote{The corresponding author is Lan Xu (xulan1@shanghaitech.edu.cn). }

\begin{abstract}

Volumetric video represents a transformative advancement in visual media, enabling users to freely navigate immersive virtual experiences and narrowing the gap between digital and real worlds.
However, the need for extensive manual intervention to stabilize mesh sequences and the generation of excessively large assets in existing workflows impedes broader adoption.
In this paper, we present a novel Gaussian-based approach, dubbed \textit{DualGS}, for real-time and high-fidelity playback of complex human performance with excellent compression ratios. Our key idea in DualGS is to separately represent motion and appearance using the corresponding skin and joint Gaussians. Such an explicit disentanglement can significantly reduce motion redundancy and enhance temporal coherence. We begin by initializing the DualGS and anchoring skin Gaussians to joint Gaussians at the first frame. Subsequently, we employ a coarse-to-fine training strategy for frame-by-frame human performance modeling. It includes a coarse alignment phase for overall motion prediction as well as a fine-grained optimization for robust tracking and high-fidelity rendering.
To integrate volumetric video seamlessly into VR environments, we efficiently compress motion using entropy encoding and appearance using codec compression coupled with a persistent codebook. Our approach achieves a compression ratio of up to 120 times, only requiring approximately 350KB of storage per frame. We demonstrate the efficacy of our representation through photo-realistic, free-view experiences on VR headsets, enabling users to immersively watch musicians in performance and feel the rhythm of the notes at the performers' fingertips. Project page: \href{https://nowheretrix.github.io/DualGS/}{https://nowheretrix.github.io/DualGS/}.

\end{abstract}

\begin{teaserfigure}
    \centering
  \includegraphics[width=\textwidth]{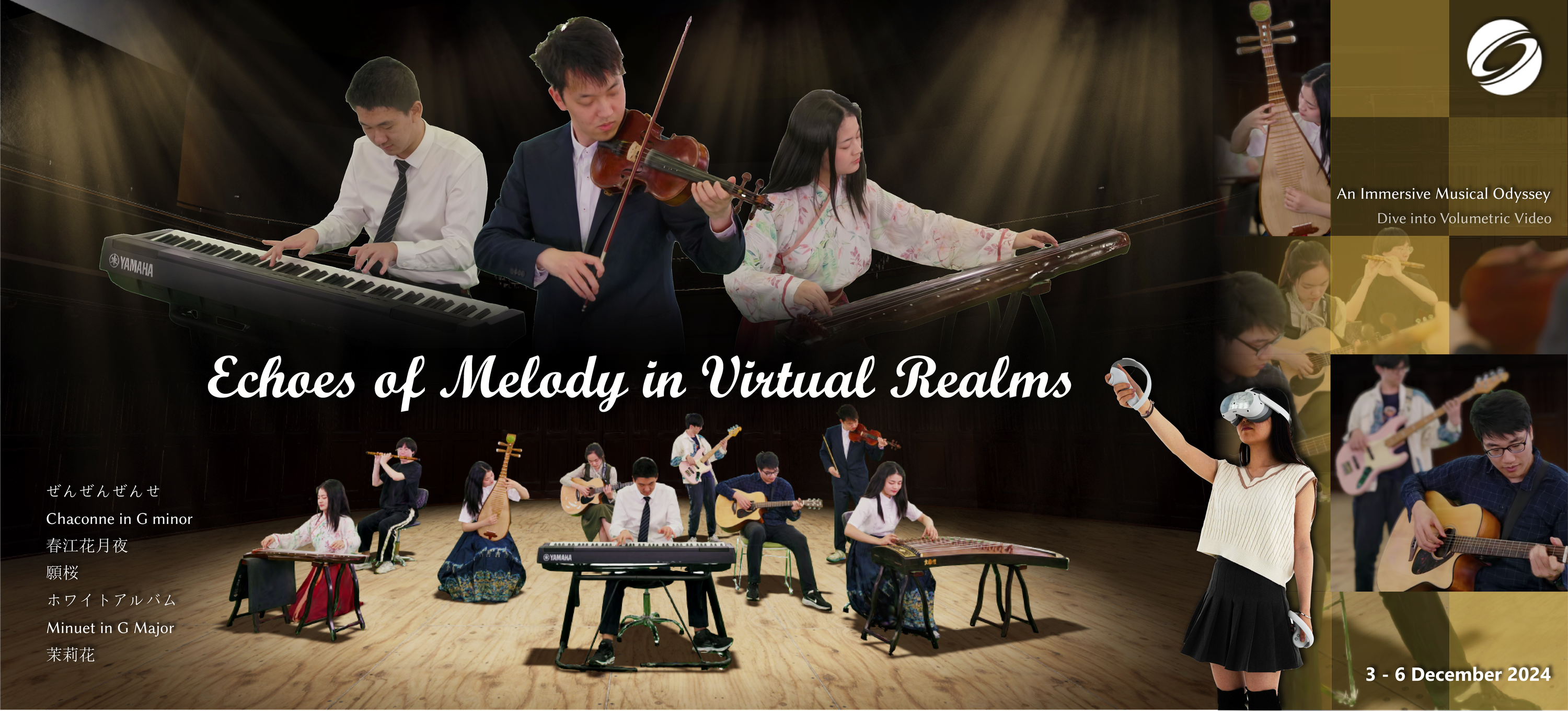}
  \vspace{-20pt}  
  \caption{We present a robust human performance tracking and rendering approach with a customized compression scheme. Our method serves as a ``ticket'' to a virtual world, offering immersive, high-fidelity viewing of multiple musicians performing.}
  \label{fig:teaser}
\end{teaserfigure}



\maketitle

\section{Introduction}
As the distinction between digital and real worlds diminishes, 3D and 4D content is rapidly gaining prominence, reshaping societal expectations and applications across digital landscapes. Among these innovations, volumetric videos represent a significant advancement in visual media. They provide viewers with six degrees of freedom, enabling users to navigate virtual environments freely. Specifically, users can immerse themselves in a virtual musical odyssey, observing musicians perform up close and feeling the rhythm of the music as if standing beside them (see Fig.~\ref{fig:teaser}).

Over the past two decades, numerous studios~\cite{vlasic2009dynamic, collet2015high, zhang2022neuvv, isik2023humanrf} have established multi-view domes worldwide, from west to east, for capturing volumetric videos. Yet, the predominant workflow for producing human-centric volumetric videos still relies on the explicit reconstruction and tracking of textured meshes. This method is prone to occlusions, often resulting in holes and noises that degrade texturing quality. Creating even minimally immersive segments requires substantial computational resources and meticulous cleanup by skilled artists. Moreover, the volumetric assets are often too large for storage and integration into immersive devices. As a result, volumetric video has not achieved widespread adoption.

Neural advances in photo-realistic rendering, notably through Neural Radiance Fields~\cite{nerf}, have facilitated bypassing explicit reconstructions and enhancing novel view synthesis.
Recently, 3D Gaussian Splatting (3DGS) advances the explicit paradigm by using learnable Gaussians to achieve high-fidelity rendering at unprecedented frame rates. It emergently facilitates the development of various dynamic variants.
For animatable avatar modeling, many works~\cite{hu2023gauhuman, kocabas2023hugs, li2024gaussianbody, pang2024ash} transform 3D Gaussians to posed space using linear blend skinning. 
For volumetric video playback, some studies~\cite{wu20234d,yang2023deformable} combine 3DGS with MLPs to model temporal coherence, sacrificing the explicit and GPU-friendly beauty of 3DGS. Yet, these methods are still fragile to challenging motions and require significant storage.

In this paper, we present a novel Gaussian-based representation for volumetric videos, achieving robust human performance tracking and high-fidelity rendering. Our core idea is to utilize Dual Gaussians, named \textit{DualGS}, for disentangled and hierarchical motion and appearance representation. It significantly enhances temporal coherence and tracking accuracy and also enables a companion compression strategy.
Our approach achieves significant storage efficiency, requiring only approximately 350KB of storage per frame. DualGS also maintains highly competitive rendering quality and consistently delivers superior rendering and temporal consistency across various challenging cases.

In DualGS, inspired by the SMPL model~\cite{SMPL2015}, which represents skin motion by interpolating a few joints, we utilize a compact number of motion-aware \textit{joint Gaussians} to capture global movements and a larger set of appearance-aware \textit{skin Gaussians} for visual representation. 
For the initialization of our DualGS representation in the first frame, we randomly initialize joint Gaussians and carefully control their scale and size to effectively represent the overall movement of the performance. Once optimized, these joint Gaussians serve as the basis for initializing the skin Gaussians. To establish the relationship between dual Gaussians, each skin Gaussian is anchored to multiple joint Gaussians, facilitating the interpolation of position and rotation for sequential optimization. 
Then, for the subsequent frame-by-frame human performance tracking, we employ a novel coarse-to-fine optimization strategy that enhances both temporal coherence and rendering fidelity.
During the coarse alignment phase, we perform optimization only on the joint Gaussians, using a locally as-rigid-as-possible regularizer while maintaining fixed appearance attributes. We also integrate a motion prediction module to aid this phase and ensure robust tracking. 
In the fine-grained optimization phase, we recompute the skin Gaussian motions from joint data as well as fine-tune the detailed positions and appearances using temporal regularizers in a differentiable manner.
Such a coarse-to-fine optimization provides explicit disentanglement of the Gaussian attributes in our DualGS, and hence significantly improves the tracking accuracy. 

Despite the advancements, integrating long-duration sequences into low-end devices like VR headsets remains challenging. Benefiting from our explicit DualGS representation, we effectively separate and compress the motion and appearance attributes. Specifically, for joint Gaussians, we employ Residual-Vector Quantization combined with entropy encoding to efficiently handle the motion attributes. For skin Gaussians, we first employ codec compression for spatial-temporal Look-up Tables, addressing both scaling and opacity attributes. Then, to manage the storage-intensive spherical harmonic (SH) attributes, we design a specialized persistent codebook. This codebook compresses SH attributes into persistent SH indices, coupled with length encoding. 
Our approach achieves a compression ratio of up to 120 times compared to the original 3DGS. It enables the seamless integration of multiple 4D assets (illustrated with 9 performers in Fig.~\ref{fig:teaser}) into VR environments for real-time rendering. This capability enables users to experience the notes pouring from the musician's dancing fingertips, embarking on a deeply immersive and enchanting musical odyssey.

\section{Related Work} 
\paragraph{Human Performance Capture.}
Recent research on human performance capture aims to achieve detailed registration for various applications~\cite{KillingFusion2017cvpr,xiang2020monoclothcap, LiveCap2019tog, Wang2021CVPR, li2021deep, zhang2023closet, zhao2022human, shao2022floren}.
Starting with the pioneering work DynamicFusion~\cite{newcombe2015dynamicfusion}, which benefits from the GPU solver to achieve real-time capture, VolumeDeform ~\cite{innmann2016volumedeform} combines depth-based correspondences with sparse SIFT features to reduce drift. 
Fusion4d~\cite{dou2016fusion4d} and Motion2fusion~\cite{motion2fusion} utilize a key-frame strategy to handle topological changes. \jyh{KillingFusion~\cite{KillingFusion2017cvpr} and SobolevFusion~\cite{slavcheva2018sobolevfusion} address these variations by introducing additional constraints on the motion fields.}
For more robust tracking, DoubleFusion~\cite{DoubleFusion} proposes a two-layer representation aided by a human parametric model, extended by UnstructureFusion~\cite{UnstructureLan} for unstructured setups. RobustFusion~\cite{robustfusion, su2022robustfusionPlus} further addresses the challenging human-object interaction scenarios. DDC~\cite{realTimeDDC} learns the deformations with skeletons and embedded graph~\cite{sumner2007embedded} and DELIFFAS~\cite{kwon2024deliffas} parameterized the light field based on DDC. Other efforts~\cite{yu2021function4d, jiang2022neuralhofusion, jiang2023instant} marry the non-rigid deformation with implicit neural advances for better performance.
Nevertheless, these methods rely on parametric template priors, focusing more on overall tracking accuracy, which limits their ability to capture fine details like wrinkles and high-frequency texture.
\paragraph{Neural Human Modeling.} In the domain of digital human neural representation, various 
approaches~\cite{lin2023im4d, xiang2022dressing, liu2020NeuralHumanRendering, lin2022efficient, suo2021neuralhumanfvv, sun2021HOI-FVV, shetty2023holoported} have been proposed to address this challenge. A collection of studies~\cite{pumarola2020d, tretschk2021nonrigid,xian2021space} model time as an additional latent variable into the NeRF's MLP. 
\jyh{For dynamic human modeling, some methods~\cite{liu2021neural, habermann2021real, habermann2023hdhumans, zhu2023trihuman, luvizon2023relightable} leverage skeleton-based and graph embedding representations, while another line of studies~\cite{shen2023x, jiang2023instantavatar, li2022tava, ARAH:ECCV:2022} build upon the SNARF~\cite{chen2021snarf} framework, which learns skinning weights through root-finding, resulting in enhanced reconstruction accuracy and improved animation quality. 
Humannerfs~\cite{zhao2022humannerf, weng_humannerf_2022_cvpr} utilize the human prior SMPL~\cite{SMPL2015} model as an anchor to warp the radiance field. NeuVV~\cite{zhang2022neuvv} and Fourier PlenOctrees~\cite{wang2022fourier} leverage advanced PlenOctree~\cite{yu2021plenoctrees} and volumetric fusion to achieve real-time rendering of dynamic scenes with significant acceleration.}
Recent methods~\cite{wang2023neus2, isik2023humanrf, song2023nerfplayer} draw inspiration from advanced framework~\cite{chen2022tensorf, muller2022instant} and incorporate explicit optimizable embeddings into the implicit representation to accelerate training times and rendering speeds.
\jyh{Building on the pioneering work of 3DGS, several dynamic variants~\cite{wu20234d, yang2023deformable, jena2023splatarmor, moreau2024human, qian20243dgs} utilize MLPs and human parametric models to establish temporal correspondences. GPS-Gaussian~\cite{zheng2023gps} develops an NVS system to regress Gaussian maps. Spacetime Gaussians~\cite{li2023spacetime} extend this approach by incorporating polynomials. ASH~\cite{pang2024ash} and Animatable Gaussians~\cite{li2024animatable} parameterize mesh positions in 2D space and infer Gaussian maps using a UNet architecture. }
GaussianAvatars~\cite{qian2023gaussianavatars, chen2023monogaussianavatar} bind Gaussians to the FLAME mesh for animation, while D3GA~\cite{zielonka2023drivable} relies on tetrahedral cages. HiFi4G~\cite{jiang2023hifi4g} leverages embedded deformation~\cite{sumner2007embedded} to accelerate training. However, most existing methods suffer from blurred results or struggle with fast motions. In contrast, our approach employs dual Gaussians coupled with a coarse-to-fine training strategy, enabling robust tracking and high-fidelity rendering.

\paragraph{Data Compression.}

Compact representation plays a pivotal role in 3D/4D reconstruction, attracting significant research interest. \jyh{For traditional animated meshes, numerous studies use PCA~\cite{alexa2000representing,vasa2007coddyac,luo2013compression} or mesh pre-segmentation~\cite{gupta2002compression, mamou2009tfan} to identify geometric parts of the human body to ensure connectivity consistency while others~\cite{mamou2009tfan, ibarria2003dynapack, luo2013compression}  predict vertex trajectories to maintain temporal coherence in vertex groups. }
For neural fields, several studies propose compact neural representations through 
CP-decomposition~\cite{chen2022tensorf}, rank reduction~\cite{tang2022compressible}, codec~\cite{Wang_2023_CVPR} and tri-planes~\cite{reiser2023merf, hu2023tri}. 
Recent works focus on the compression of 3D Gaussian representations. \jyh{Compact3D~\cite{navaneet2023compact3d}, C3DGS~\cite{Niedermayr_2024_CVPR} and Compact-3DGS~\cite{Lee_2024_CVPR} use vector quantization and entropy encoding while LightGaussian~\cite{fan2023lightgaussian} prunes Gaussians and adopts octree-based compression for positions. RDO-Gaussian~\cite{wang2024endtoendratedistortionoptimized3d} and Reduced3DGS~\cite{reduced3DGS} combine redundant Gaussian culling with vector quantization, whereas SOG~\cite{morgenstern2023compact} maps Gaussian attributes onto 2D grids and utilizes image codec compression techniques. Scaffold-GS~\cite{scaffoldgs} leverages anchor points to significantly reduce the number of redundant Gaussians, while HAC~\cite{chen2024hac} further enhances compression with a combination of hash tables and learnable features. EAGLES~\cite{girish2023eagles} compresses attributes using quantized latent codes and a trainable decoder. }
In the dynamic domain, 4K4D~\cite{xu20234k4d} employs a 4D feature grid and IBR module, VideoRF~\cite{wang2023videorf} encodes 4D radiance fields into 2D feature streams, \jyh{TeTriRF~\cite{wu2024tetrirf} bakes density grid sequences into the tri-plane representation,} while HiFi4G~\cite{jiang2023hifi4g} utilizes residual computation and entropy encoding. However, these methods either fail to achieve high compression ratios or compromise quality. In contrast, our method requires only 350KB per frame while maintaining high-fidelity rendering.

\begin{figure*}[t] 
	\begin{center} 
		\includegraphics[width=\linewidth]{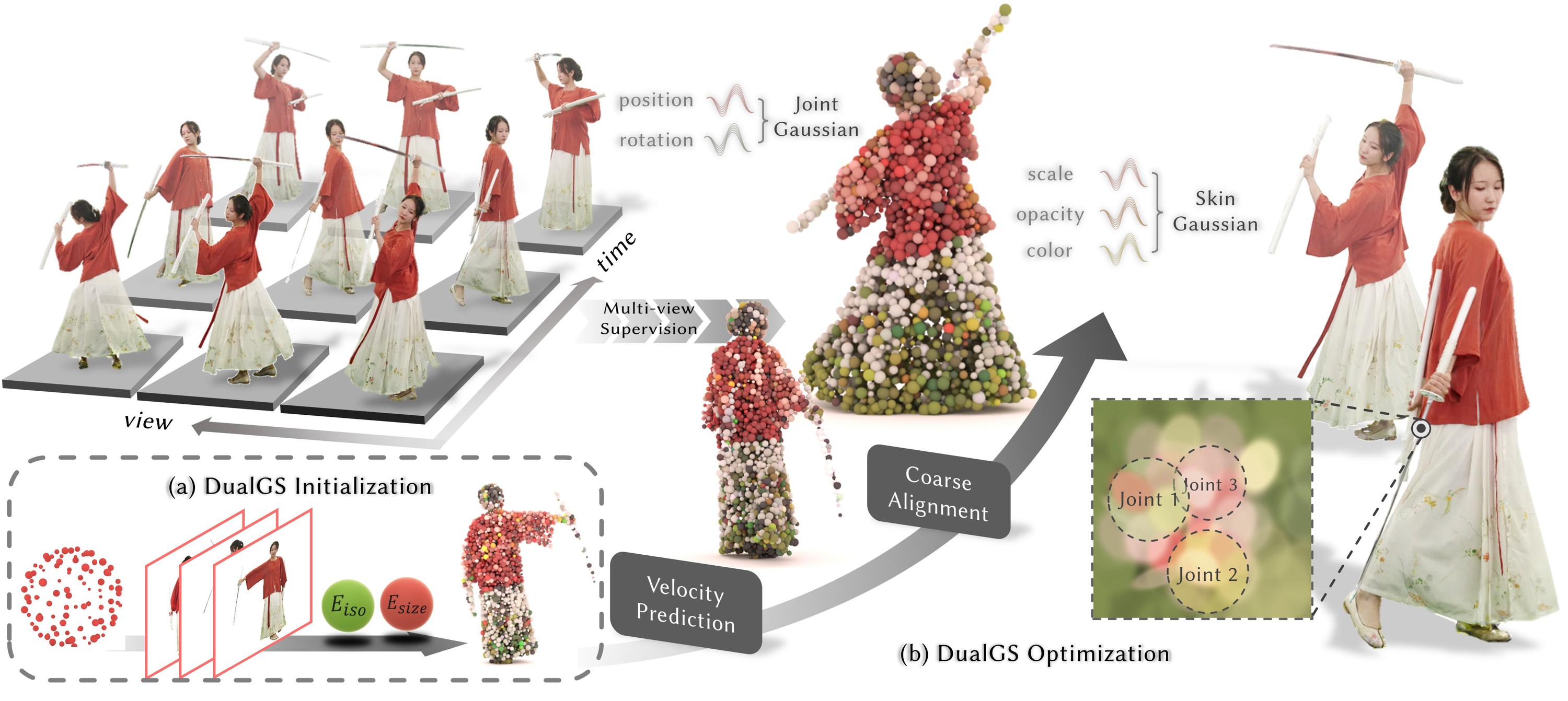} 
	\end{center} 
    \vspace{-10pt}  
    \caption{We propose a novel Dual Gaussian representation to capture challenging human performance from multi-view inputs. We first optimize joint Gaussians from a random point cloud, then use them to initialize skin Gaussians, expressing their motion through interpolation. In the following optimization, we employ a coarse-to-fine strategy, with a coarse alignment for overall motion prediction and fine-grained optimization for robust tracking and high-fidelity rendering.}
    \label{fig:overview} 
    \vspace{-8pt}
\end{figure*}

\section{Dual-Gaussian Representation}\label{sec:algorithm} 

Given multi-view videos capturing a dynamic 3D scene, our objective is to robustly track human performance and achieve high-quality novel view rendering in real-time. 
The methodology is visually summarized in Fig.~\ref{fig:overview}. We first introduce a Dual Gaussian(DualGS) representation, which comprises a small number of motion-aware joint Gaussians to capture global movements and a large set of appearance-aware skin Gaussians to express visual appearance. Additionally, we propose a novel coarse-to-fine optimization strategy with a motion prediction module to ensure temporal consistency and produce high-fidelity Gaussian assets. Our method enables accurate tracking and realistic rendering at 4K resolution, outperforming existing approaches in performance and quality.

\subsection{Dual-Gaussian Initialization} \label{sec:init}
We first initialize the DualGS to establish the mapping between the skin Gaussians and joint Gaussians.
To simplify the description, we first categorize the attributes of Gaussians into two groups:
1) motion-aware parameters, which include position $p$ and rotation $q$. 2) appearance-aware parameters, comprising spherical harmonic  $\mathcal{C}$, opacity $\sigma$, and scaling $s$.
Inspired by the human parametric model, where skin vertices are represented through the interpolation of a minimal number of joints,
our approach utilizes a dual Gaussian scheme to separately encode motion and appearance.
We utilize a compact set of Gaussians($\sim$15,000) to encapsulate the overall motion of the performer, while a more extensive set of Gaussians($\sim$180,000) captures the nuanced appearance details. 
Once established, the number of Gaussians remains constant over time, with only the attributes subject to continuous updates. 
This formulation yields two main benefits: 1) The relative positions of the skin Gaussians in local space maintain stability, driven consistently by the same joint Gaussians motion, thereby enhancing spatial-temporal consistency. 2) The reduced motion parameters are highly conducive to subsequent compression processes.

\begin{figure}[t] 
	\begin{center} 
		\includegraphics[width=0.99\linewidth]{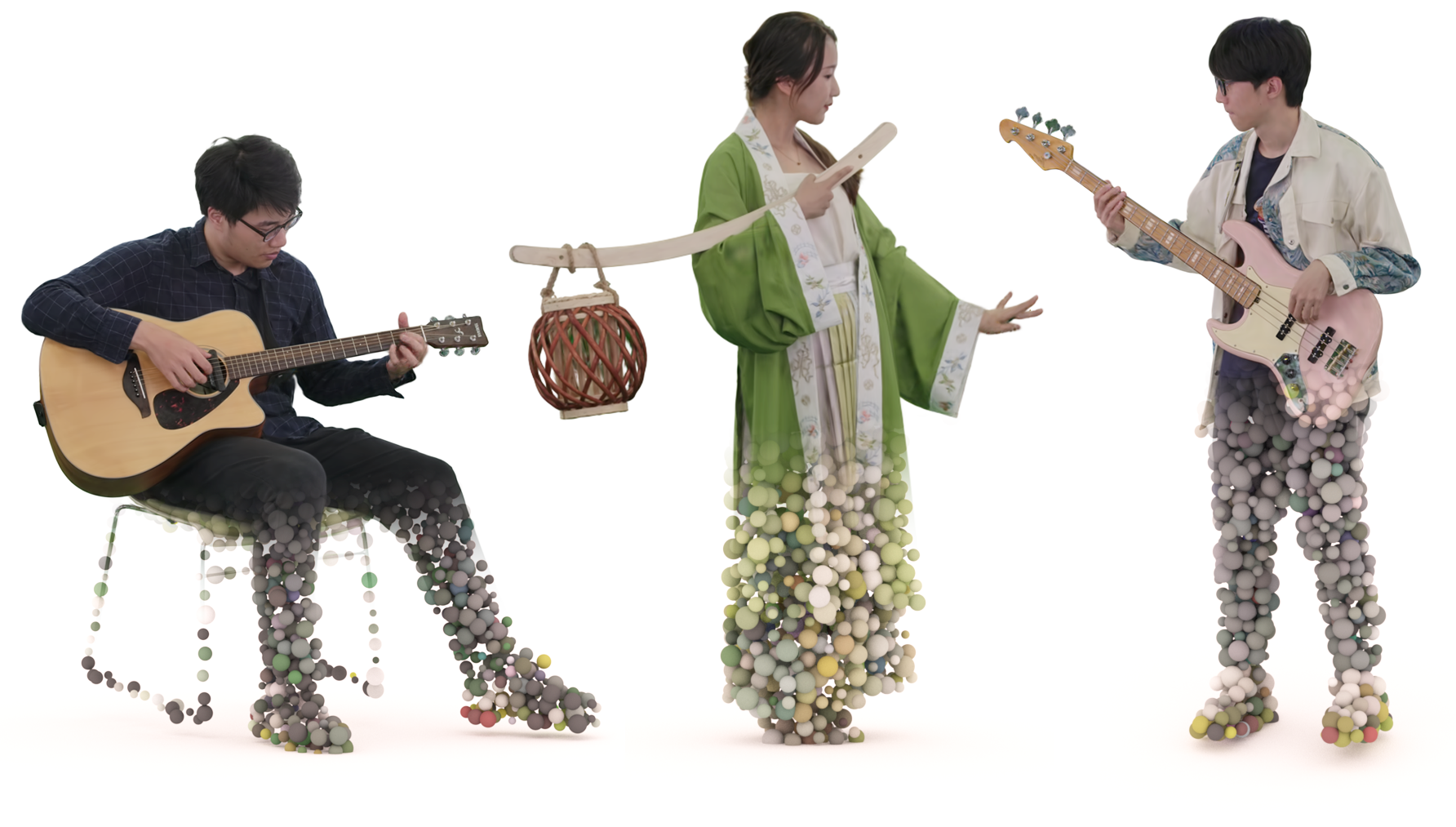} 
	\end{center} 
	\vspace{-18pt}
	\caption{\jyh{Sampled results from our DualGS optimization pipeline. With the aid of our coarse-to-fine training strategy, we can produce high-fidelity 4D assets.}}
	\label{fig:optimization_result}
 	\vspace{-5pt}

\end{figure} 

Analogous to the original 3DGS~\cite{kerbl20233d}, we commence by initializing a small number of joint Gaussians to represent global motion dynamics. \jyh{The Gaussian model is trained on the first frame using a uniform random initialization.} During the training process, we regulate the number of Gaussians to strike an optimal balance between efficient motion representation and compact storage. \jyh{Specifically, we perform densification and pruning before 15,000 iterations. The joint Gaussians are then downsampled to approximately 15,000, fixing this number and subsequently optimizing only their values.}
Skinny kernels are generated to effectively fit local appearance details but lack geometric information, leading to unexpected plush artifacts in the following optimizations. To address this, we follow PhysGaussian~\cite{xie2023physgaussian} that employs an isotropic loss that constrains overly skinny scaling:
\begin{equation}
\begin{split}
E_{\mathrm{iso}} & =\frac{1}{N} \sum_{i=1}^{N} \text{ReLU}(e^{max(s_i) - min(s_i)} - r),
\end{split}
\end{equation}
where $s_i$ represents the $i$-th joint Gaussian scaling parameters, and $e$ is the activation function. We enforce a constraint that the ratio between the length of the major and minor axis does not exceed $r$. Additionally, we propose another term to constrain oversized Gaussians, preventing local over-reconstruction:
\begin{equation}
\begin{split}
E_{\text {size }}=\sum_{i=1}^N \text{ReLU} \left(s_i- \alpha \frac{1}{N} \sum_{i=1}^N sg[s_i] \right), 
\end{split}
\end{equation}
where $sg$ stands for the stop-gradient operator. $E_{\text {size }}$ penalizes those Gaussians whose scale exceeds the average size by a factor of $\alpha$. 

For skin Gaussians initialization, we use the initialized joint Gaussians kernels as inputs and perform training to achieve high-fidelity quality. During the training process, the position of skin Gaussians is updated through differentiable rasterization. According to human parametric models~\cite{SMPL2015, li2017learning} where skin vertices are driven by predefined joint motions and skinning weights, we then bind each skin Gaussian to the k-nearest joint Gaussians. 
Specifically, for each skin Gaussian position $p^{s}_i$, we identify the k-nearest neighbors(KNN, k = 8) joint Gaussians $p^{j}_k$ within the encompassing ellipsoid to serve as its anchor joints.\jyh{ The blending weight is defined as:
\begin{equation}
w\left(p^s_i, p^j_k\right)=\exp \left(-\left\|p^s_i-p^j_k\right\|_2^2 / l^2\right),
\label{eq:blending_weight}
\end{equation}
where $l$ is the influence radius. Here, the superscripts $j$ and $s$ denote the joint Gaussians and skin Gaussians, respectively. This KNN graph and the corresponding blending weights are integral to the subsequent optimization process and remain fixed throughout.}

Notably, our approach offers greater flexibility compared to the parametric model, which relies on predefined joints, skinning weights, or a fixed topology. Experimental results demonstrate that our method can handle a wide range of dynamic sequences.

\noindent{\bf Implementation.}
We perform 30,000 training iterations for DualGS initialization separately. For joint Gaussians, the complete loss function is as follows:
\begin{equation}
E_{\text {init }} = \lambda_{iso} E_{\text {iso }} + \lambda_{size} E_{\text {size }} + E_{\text {color}}, 
\end{equation}
where $E_{\text {color}}$ is the photometric loss. We use the following empirically determined parameters: $ r= 4, \alpha = 3, l = 0.001, \lambda_{iso} = 0.005, \lambda_{size} = 1$.

\subsection{Dual-Gaussian Optimization} \label{sec:opt}
For sequential training, we fix the number of DualGS and optimize the motion of joint Gaussians as well as the appearance of skin Gaussians. We observe that Gaussians tend to alter appearance rather than update positions to the desired location to fit the photometric loss. To address this, we adopt a coarse-to-fine training strategy that starts with isolated coarse alignment and advances to integrated fine-grained optimization to achieve robust human performance tracking and high-fidelity rendering.

\paragraph{Coarse Alignment.}
Upon initializing the joint Gaussians on the first frame, we fix the color, opacity, and scaling attributes to concentrate on capturing the human dynamic motions. In this phase, we solely fine-tune the motions of the joint Gaussians. Inspired by dynamic 3d Gaussian~\cite{luiten2023dynamic}, we employ a smooth regularizer to constrain the joint Gaussians motion locally as-rigid-as-possible(ARAP):
\begin{equation}
\begin{aligned}   
E_{\text {smooth }}= & \sum_i \sum_{k \in \mathcal{N}(i)} w_{i, k} \| R\left(q^j_{i, t} * {q^j_{i, t-1}}^{-1}\right) \\
& \left(p^j_{k, t-1}-p^j_{i, t-1}\right)-\left(p^j_{k, t}-p^j_{i, t}\right) \|_2^2,
\end{aligned}
\end{equation}
where $\mathcal{N}(i)$ represents the set of neighboring joint Gaussian kernels of $i$, $R(\cdot)$ converts quaternion back into a rotation matrix and \jyh{$w_{i, k}$ corresponds to the blending weights defined in Eq.~\ref{eq:blending_weight}. These weights remain fixed throughout the optimization process to avoid additional storage overhead. 
Additionally, following the original 3DGS~\cite{kerbl20233d}, we incorporate the $\mathcal{L}_1$ photometric loss combined with a D-SSIM term during the coarse alignment process. The color energy is defined as:
\begin{equation}
E_{\text {color }}=(1-\lambda_{color}) \mathcal{L}_1+\lambda_{color} \mathcal{L}_{\text {D-SSIM}},
\end{equation}
the complete energy for coarse alignment is as follows:
\begin{equation}
E_{\text {coarse}} = \lambda_{smooth}^j E_{\text {smooth }} + E_{color}^j, 
\end{equation}
}
where $E_{color}^j$ is computed by comparing the blended color after joint Gaussians rasterization with the ground truth input images.

\paragraph{ Motion Prediction.} To handle challenging motions, we further maintain a velocity attribute for each Gaussian and use the position changes between the latest two frames for weighted updates. Before the coarse alignment, we first estimate the new frame Gaussian positions based on the last one and velocities, then apply non-rigid constraint(ARAP) to restrict the unreasonable motions.
\begin{figure}[tbp] 
	\centering 
	\includegraphics[width=1\linewidth]{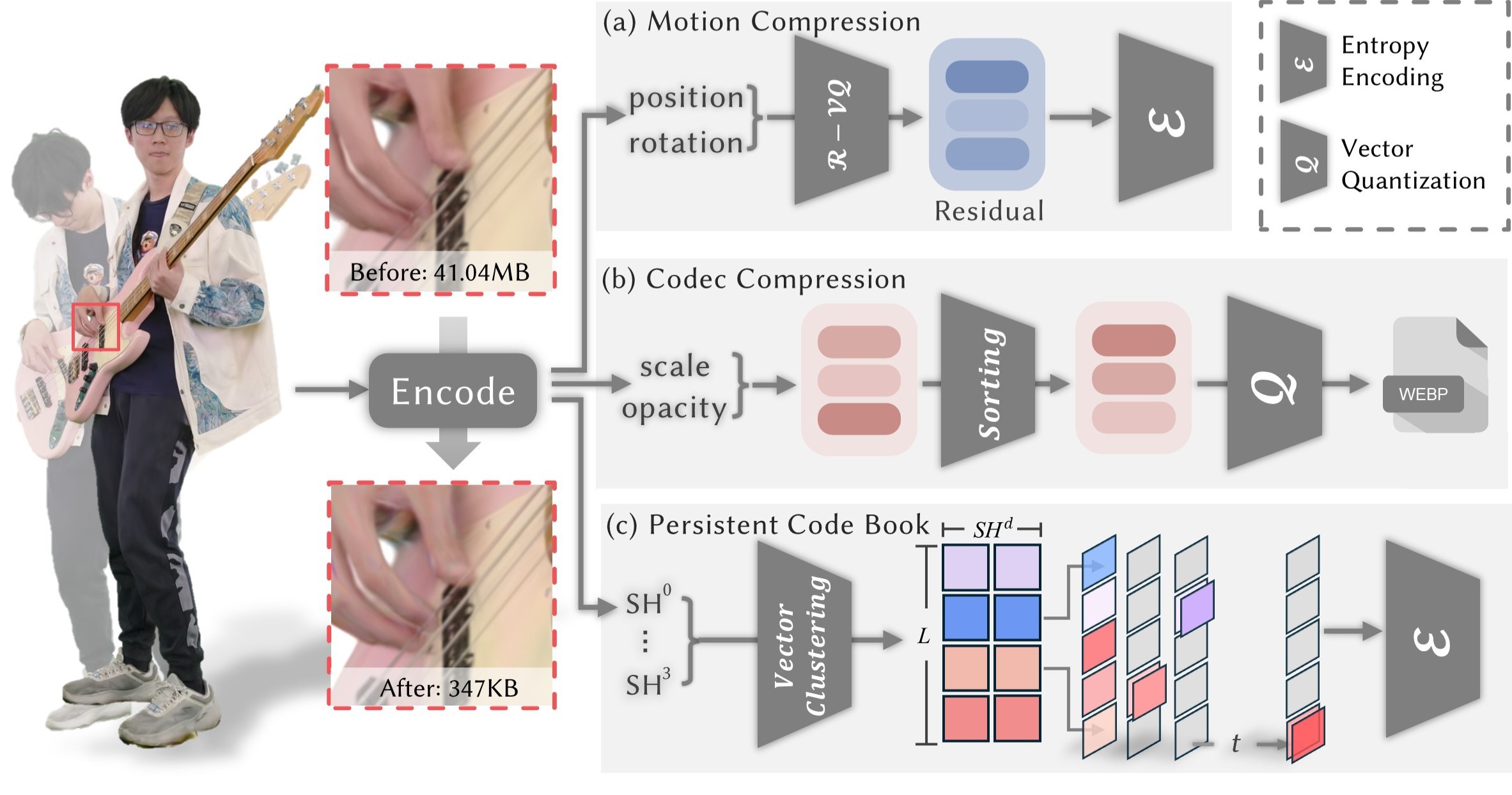} 
	\vspace{-15pt} 
	\caption{Illustration of our hybrid compression strategy. We compress joint Gaussian motions using residual vector quantization, encode opacity and scaling via codec compression, and represent spherical harmonics with a persistent codebook. Our approach achieves a compression ratio of up to 120-fold.} 
	\label{fig:compress} 
	\vspace{-8pt} 
\end{figure} 
\paragraph{Fine-grained Optimization.}
We then optimize the motion of joint Gaussians and the appearance of skin Gaussians via the differentiable tracking and rendering process. Using the joint Gaussians motion from the coarse alignment phase, we interpolate the position and rotation of the skin Gaussians, balancing the rendering quality and temporal consistency:
\begin{equation}
\begin{aligned}
{q}_{i, t}^s= & \sum_{k \in \mathcal{N}\left(p_{i,1}^s\right)} w\left(p_{i, 1}^s, p_{k, 1}^j\right)  q_{k, t}^j, \\
{p}_{i, t}^s= & \sum_{k \in \mathcal{N}\left(p_{i,1}^s\right)} w\left(p_{i, 1}^s, p_{k, 1}^j\right) (R(q_{k, t}^j)p_{i,1}^s + p_{k,t}^j), 
\end{aligned}
\end{equation}
where $\mathcal{N}\left(p_{i,1}^s\right)$ and $w\left(p_{i, 1}^s, p_{k, 1}^j\right)$ represent the precomputed KNN graph and blending weights from the initialization stage respectively. During backpropagation, the gradients on the skin Gaussians  ${q}_{i, t}^s, {p}_{i, t}^s$ are further propagated along the computation graph to the joint Gaussians $q_{k, t}^j$ and $p_{k,t}^j$.
\jyh{Furthermore, to enhance temporal consistency, we incorporate a temporal regularization term inspired by HiFi4G~\cite{jiang2023hifi4g}. This term constrains the 4D Gaussian appearance attributes $\left(\mathcal{C}_{i, t}, \sigma_{i, t}, s_{i, t}\right)$ from undergoing significant updates between consecutive frames:
\begin{equation}
E_{\text {temp }}=\sum_{a \in\{\mathcal{C}, \sigma, s\}} \lambda_{\mathrm{a}}\left\|a_{i, t}-a_{i, t-1}\right\|_2^2, 
\end{equation}
$E_{\text {temp }}$ efficiently improves the visual quality while enabling higher compression ratios in the subsequent later stage.} We define the overall energy as follows:
\begin{equation}
\begin{aligned}
E_{\text {fine }} = \lambda_{smooth}^s E_{\text {smooth}} + \lambda_{temp} E_{\text {temp}} + E^s_{\text {color}},
\end{aligned}
\end{equation}
\jyh{We visualize the joint Gaussian kernels and the corresponding skin Gaussians rendering in Fig.~\ref{fig:optimization_result}. With the aid of our coarse-to-fine training strategy, DualGS efficiently achieves robust human performance tracking and high-fidelity rendering.}
Regarding our implementation, we first employ velocity prediction to initialize the motions, then conduct 10,000 iterations of training in each phase. The hyperparameters are set as follows: $\lambda_{color} = 0.2$, $\lambda_{smooth}^j = 0.05, \lambda_{smooth}^s = 0.001, \lambda_{\mathcal{C}} = 1, \lambda_{\sigma} = 0.003,   \lambda_{s}=0.003, \lambda_{temp} = 0.00003$.

\begin{figure}[t] 
	\centering 
	\includegraphics[width=1\linewidth]{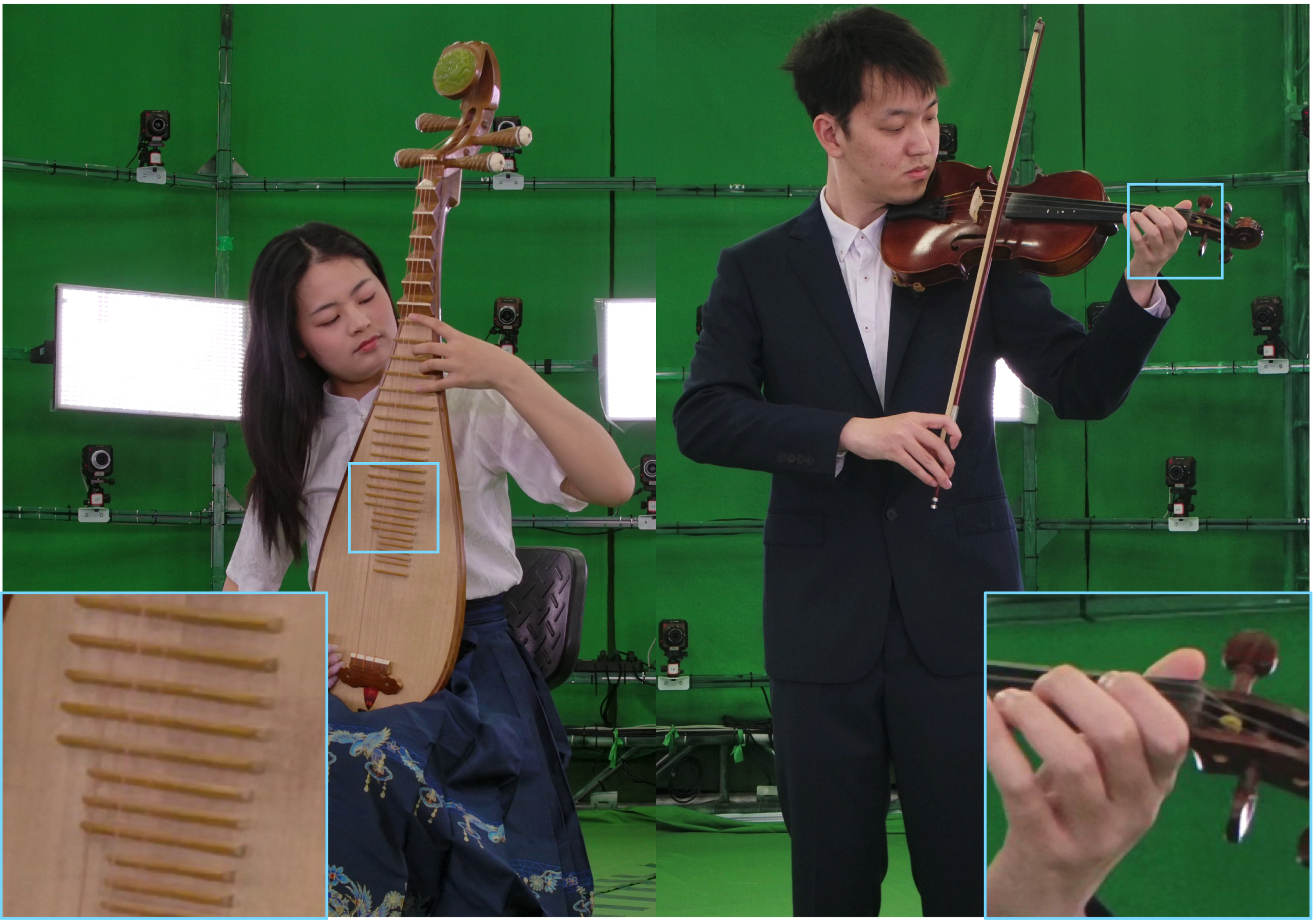} 
	\vspace{-20pt} 
	\caption{\jyh{Examples of data captured by our multi-view system. Our DualGS dataset includes a diverse range of musical instruments from both Western and Eastern traditions.}} 
	\label{fig:dataset} 
	\vspace{-8pt} 
\end{figure}

\section{Compression}\label{sec:compression}

Our goal is to seamlessly integrate the high-quality 4D assets generated by DualGS into low-end devices with limited memory, such as head-mounted displays. For example, users can immersively navigate a musical odyssey in a VR environment. However, integrating such multiple volumetric videos (9 people in Fig.~\ref{fig:teaser}) totaling 2700 frames is non-trivial, requiring over 130GB of storage and even more runtime memory. Thanks to the effective disentanglement provided by DualGS, we organically compress the motion and appearance separately from joint Gaussians and skin Gaussians. Our strategy achieves a compression ratio of up to 120 times, while still enabling the decoding of high-fidelity rendering results in real-time. 
We first divide the sequences into multiple segments, with each segment consisting of $f$ frames(50 in our setting).

\begin{figure*}[t] 
	\begin{center} 
		\includegraphics[width=1.0\linewidth]{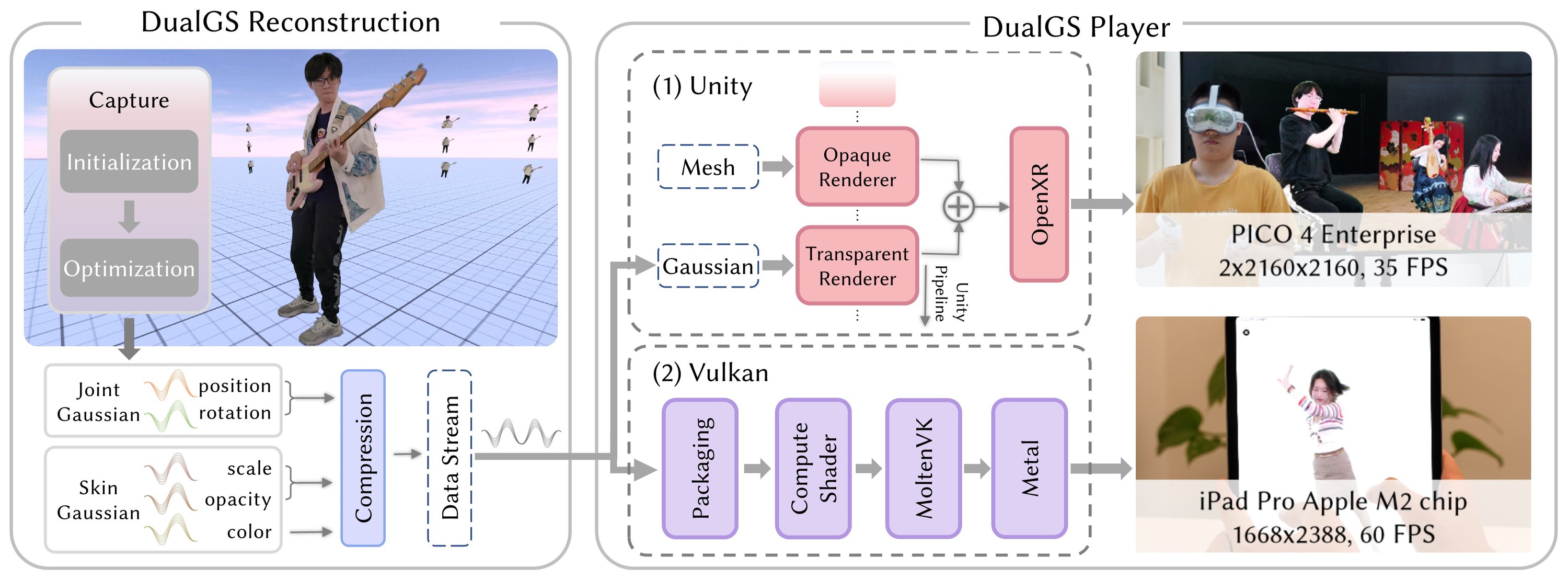} 
	\end{center} 
	\vspace{-13pt}
	\caption{\jyh{Illustration of our DualGS player implementation for the seamless integration of 4D sequences into Unity and mobile platforms, enhancing real-time immersive rendering across multiple devices.}} 
	\label{fig:application}
	\vspace{-15pt}
\end{figure*}

\paragraph{Residual Compression.}
As mentioned in C3DGS~\cite{niedermayr2023compressed} and CompGS~\cite{liu2024compgs}, the precision of Gaussian position plays a crucial role in the quality of the scene, where even minor errors can severely impact rendering quality. Therefore, they opt for high-bit quantization. To address this, we first employ Residual-Vector Quantization(R-VQ) on the joint Gaussians motion. We retain the position of the first frame $R_{i,1} = p_{i,1}^j$ in the current segment, then perform temporal quantization(11-bit in our setting) as follows:
\begin{equation}
R_{i,t}= Q\left(p_{i,t}^j-(R_{i,1} + \sum_{k=2}^{t-1} Q^{-1}\left(R_{i,k}\right))\right), t>1
\end{equation}
$Q$ and $Q^{-1}$ represent the quantization and dequantization respectively. We also apply R-VQ to the rotation $q$. Compared to solely quantizing adjacent frame residuals, our scheme effectively prevents error accumulation. We further employ Ranged Arithmetic Numerical System(RANS) encoding for lossless compression.

\paragraph{Codec Compression.} Although we can apply residual compression to opacity and scaling parameters, the significantly larger number of skin Gaussians results in a notably higher storage requirement for compressed opacity and scaling. 
To achieve an optimal balance between data accuracy and storage overhead, we leverage the spatial-temporal relationships of these two Gaussian attributes. By benefiting from the temporal regularizer, we embed the opacity and scaling into separate Look-up Tables (LUT) and then apply image codec compression for encoding. Specifically, the opacity and scaling attributes are arranged into a 2D LUT, with the height corresponding to the number of skin Gaussians and the width corresponding to the segment frame length. To enhance 2D consistency, we further sort the LUT by the average value of each row. We then quantize and compress the 2D LUT using an image codec(WebP/JPEG), encoding it as an 8-bit image with a quality level of 100.

\paragraph{Persistent Code Book.}
The color attributes take up the majority of storage, occupying 48 out of 59 parameters. Effectively compressing them can yield significant storage savings. However, applying residual or codec compression to these coefficients still requires considerable storage overhead. 
To this end, we design a novel compression strategy -- a persistent codebook that leverages the temporal consistency of skin Gaussian SH parameters, achieving up to a 360-fold compression. 
In particular, we apply K-Means clustering to the d-order($d=0,1,2,3$) SH coefficients across all frames within this segment. The codebook $\mathcal{Z}_d$ is initialized with a uniform distribution and iteratively updated by randomly selecting a batch of $d$-order coefficients. After optimization, we obtain four codebooks of length $L$(8192 in our setting). 
The skin Gaussian SH attributes are compactly encoded to SH indices via these codebooks:
\begin{equation}
\tau_{i,t}^d =\underset{k \in \{1,\ldots,L\}}{\operatorname{argmin}}\left\|\mathcal{Z}_d[k]-\mathcal{C}_{i,t}^d\right\|_2^2, 
\end{equation}
where $\tau_{i,t}^d$ is the $d$-order SH index for skin Gaussian $i$ at frame $t$. 
We also can recover the compressed SH parameters $\hat{\mathcal{C}}_{i,t}^d$ by indexing into the codebooks $\hat{\mathcal{C}}_{i,t}^d = \mathcal{Z}_d[\tau_{i,t}^d]$ .
Using this representation, the SH attributes, originally consisting of $n \times f \times 48$ float parameters, are encoded as $n \times f \times 4$ integer indices along with four distinct codebooks, where $n$ is the number of skin Gaussian. Furthermore, we observe that temporally coherent SH coefficients still maintain high consistency after being converted into indices. According to our calculation, on average, only one percent of the skin Gaussians SH indices change between adjacent frames. Therefore, instead of saving the spatial-temporal SH indices for each frame, we only save the first frame indices and the positions where the indices change in adjacent frames. Specifically, if $\tau_{i,t}^d\neq\tau_{i,t-1}^d$, we update it to the new index $\tau_{i,t}^d=k$ and save this integer quadruples $(t, d, i, k)$. In real-time decoding, we can instantly decode the spatial-temporal SH indices for each frame based on these quadruples. Additionally, the order of the quadruples does not affect the decoding process. We sort them in ascending order based on the first two variables and then apply length encoding.

\begin{figure*}[htbp] 
	\begin{center} 
		\includegraphics[width=1\linewidth]{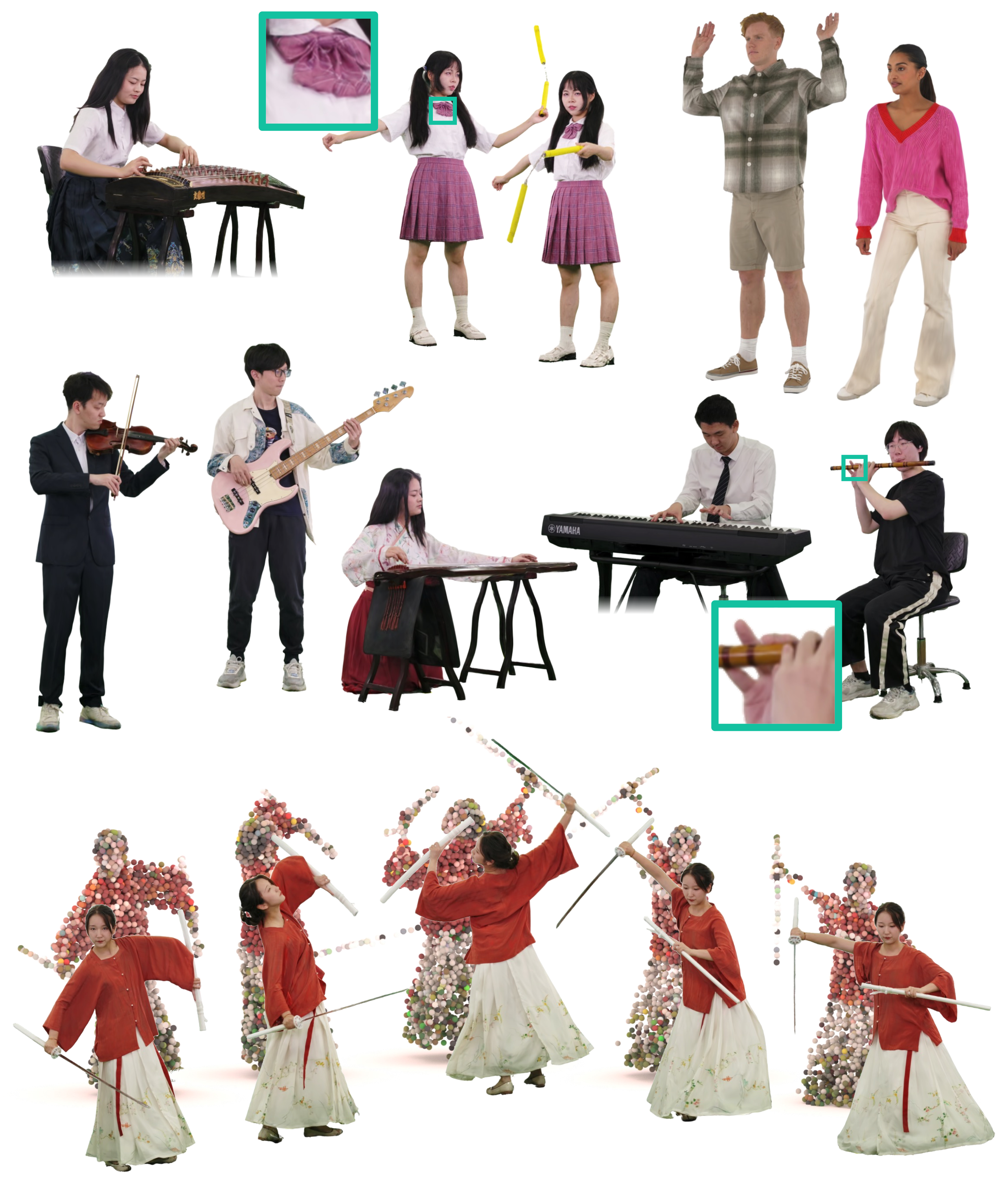} 
	\end{center} 
	\vspace{-13pt}
	\caption{We present a comprehensive results gallery showcasing our robust Dual Gaussian Splatting pipeline, featuring complex scenarios such as nunchuck swinging, musical instrument playing, and dancing. Additionally, we visualize dynamic sequences along with the corresponding joint Gaussians tracking. Even in the presence of challenging motions, DualGS achieves a 120-fold compression while maintaining real-time, high-fidelity rendering of human performances.} 
	\label{fig:gallery}
	\vspace{-15pt}
\end{figure*}

\begin{figure*}[htbp] 
	\begin{center} 
		\includegraphics[width=0.99\linewidth]{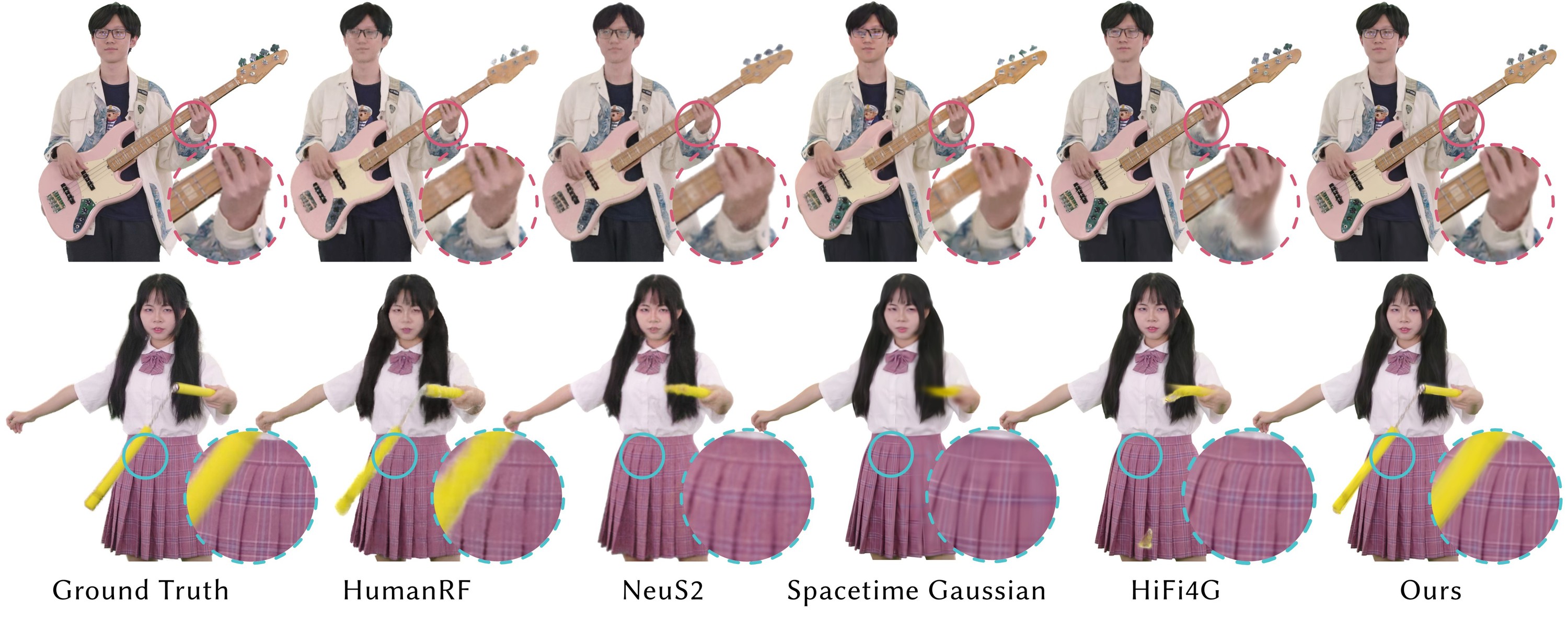} 
	\end{center} 
	\vspace{-16pt}
	\caption{Qualitative comparison of our method against HumanRF~\cite{isik2023humanrf}, NeuS2~\cite{wang2023neus2}, Spacetime Gaussian~\cite{li2023spacetime} and HiFi4G~\cite{jiang2023hifi4g} on our challenging dataset. Our method achieves the highest rendering quality.}
	\label{fig:fig_comp_1}
 	\vspace{-9pt}

\end{figure*} 
\jyh{
\section{Implementation}\label{sec:Implementation} 

\subsection{Dataset and Training Details} 

\label{sec:dataset}
We utilize 81 Z-CAM cinema cameras to capture challenging human performances with a resolution of 3840 $\times$ 2160 at 30 fps under global illumination. To minimize motion blur during fast actions, all cameras are configured with a shutter speed of 640 µs, ensuring crisp and clear video quality. We showcase data examples in  Fig.~\ref{fig:dataset}.
Our DualGS dataset features 8 actors performing a wide range of musical instruments from both Western and Eastern traditions, such as violin, guitar, piano, flute, lute, and guzheng. Each sequence in the dataset starts with a standard pose to mitigate the close-to-open issue.
These performances span various styles — from graceful classical melodies to contemporary pop music and vibrant subcultural pieces, providing a detailed portrayal of the
performers’ nimble finger movements and expressive demeanors.
Additionally, for each instrument, the performers play pieces with slow, medium, and fast tempos, allowing us to demonstrate the robustness of our method across different motions. As illustrated in Fig.~\ref{fig:gallery}, DualGS enables robust tracking and high-fidelity rendering of human-centric volumetric video in real-time at high resolutions.  
For data pre-processing, we apply the background matting~\cite{BGMv2} to extract the foreground masks from all captured frames.
Regarding DualGS optimization, due to limited GPU memory, we employ the per-frame training strategy. In the coarse alignment phase, we use the same learning rate as 3DGS~\cite{kerbl20233d}. For the Fine-grained phase, we reduce the learning rate and schedules by a factor of 10, resetting them at the beginning of each frame. We train the multi-view sequences on a single NVIDIA GeForce RTX3090, achieving a processing time of 12 minutes per frame.
For rendering, DualGS adds an extra 10 ms for attribute decoding, memory copying, and skin Gaussians motion interpolation, achieving 77 fps for 4K rendering. Notably, the decoding process leverages CPU resources, running in parallel with Gaussian rasterization for acceleration.

\subsection{DualGS player} \label{sec:player}

As illustrated in Fig.~\ref{fig:application}, for the compressed data stream, we develop a companion Unity plugin that seamlessly integrates long-duration 4d sequences into standard CG engines and VR headsets, allowing conventional 3D rendering pipelines to efficiently deliver immersive environments. Additionally, we implement a DualGS player that enables real-time rendering on low-end mobile devices, offering a more user-friendly and interactive experience. 
\paragraph{Unity Plugin.}
Based on the open-source Unity Renderer~\cite{Unity3DGS}, we implement a rendering plugin in Unity based on OpenXR that not only supports importing 4D assets generated by DualGS into Unity but also addresses shading and occlusions, seamlessly fusing the environment with Gaussian rasterization results in delivering an immersive, high-fidelity experience.
Firstly, we decode the Gaussian point cloud of the current frame from the compressed data stream. Following the differentiable rasterization, the rendered images and corresponding alpha channels are stored in an additional texture buffer. This texture buffer is then combined with those produced by the standard mesh rendering pipeline, performing alpha blending from back to front to correctly handle occlusions. Leveraging Unity's cross-platform capabilities, we can further stream the content to VR headsets, allowing users to experience immersive viewing in the virtual environment.

\paragraph{Gaussian Renderer.}

To enable high-fidelity dynamic rendering on low-end devices like iPhones and iPads, we developed a companion rendering application based on Vulkan~\cite{3DGScpp}, removing the reliance on high-end GPU hardware. Our compression strategy explicitly divides sequences into multiple segments (50 frames per segment), allowing seamless playback of any length. Specifically, we employ a multi-threaded approach to parallelize data loading, decoding, and rendering. Once a frame is decoded, the Gaussian kernels are packaged into storage buffers compatible with compute shaders, which are then rendered directly to the swapchain image using alpha blending.
The application is compatible with multiple platforms, including Windows, Linux, and Android. To extend support to iOS devices, such as Vision Pro, iPhones, and iPads, we leverage the MoltenVK library to map Vulkan API calls to Metal API. Within our DualGS player, users can drag, rotate, pause, and play the volumetric video, enhancing both accessibility and versatility across a wide range of devices.
}

\section{Experiments}

\subsection{Comparison} 

\begin{table}[t]
	\begin{center}
		\centering
		\vspace{-1pt}
        \caption{
        \textbf{Quantitative comparison with SOTA dynamic rendering methods on our DualGS dataset}. Green and yellow cell colors indicate the best and the second-best results.}
        \vspace{-2pt}
		\resizebox{0.47\textwidth}{!}{
			\begin{tabular}{l|ccccc}
				\hline
				Method   &  PSNR $\uparrow$ & SSIM $\uparrow$ & LPIPS $\downarrow$ & VMAF$\uparrow$ & Storage(MB / frame) $\downarrow$ \\
				\hline
			    HumanRF~\cite{isik2023humanrf}     & 29.701 & 0.969 & 0.0461 & 79.171 & 7.566 \\
				NeuS2~\cite{wang2023neus2}\qquad\qquad & 29.417 & 0.970 &  0.0593 & 77.912 & 24.163 \\
                Spacetime Gaussian~\cite{li2023spacetime} & 29.532 & 0.964 & 0.0362 & 70.923 & \colorbox{best2}{0.846} \\
				HiFi4G~\cite{jiang2023hifi4g}  & 33.503 & 0.988 & 0.0239 & 84.737 & 1.581 \\
    		\hline
                Ours(Before Compression)  & \colorbox{best1}{35.577} & \colorbox{best1}{0.990} & \colorbox{best1}{0.0196} & \colorbox{best1}{86.504} & 42.020\\
                Ours(After Compression) & \colorbox{best2}{35.243} & \colorbox{best2}{0.989} & \colorbox{best2}{0.0221} & \colorbox{best2}{86.171} & \colorbox{best1}{0.323}\\
				\hline
			\end{tabular}
		}
		\label{table:comparison1}
    
    	\vspace{-3pt}
	\end{center}
\end{table}

\paragraph{Rendering Comparison.} We compare DualGS against SOTA implicit Instant NGP-based methods, HumanRF~\cite{isik2023humanrf} and NeuS2~\cite{wang2023neus2} as well as explicit Gaussian-based methods Spacetime Gaussian~\cite{li2023spacetime} and HiFi4G~\cite{jiang2023hifi4g} using our captured dataset. As illustrated in Fig.~\ref{fig:fig_comp_1}, HumanRF~\cite{isik2023humanrf} produces blurry results, whereas NeuS2~\cite{wang2023neus2} struggles with high-frequency details. Spacetime Gaussian~\cite{li2023spacetime} is prone to oversmoothing, losing fine details such as clothing wrinkles, while HiFi4G~\cite{jiang2023hifi4g} heavily relying on explicit mesh reconstruction and non-rigid tracking, generates severely unnatural results where deformation fails. Additionally, these methods exhibit artifacts or fail in rapid-motion areas. In contrast, our template-free DualGS leverages a dual Gaussian representation coupled with a tailored compression scheme for precise tracking and high-fidelity rendering.
Our approach not only produces GPU-friendly and memory-efficient 4D assets but also demonstrably outperforms the compared methods. 
For quantitative comparison, we evaluate each method across three sequences, each consisting of 200 frames. In addition to traditional metrics such as PSNR, SSIM, and LPIPS, we introduce per-frame storage and VMAF~\cite{2016Toward} to evaluate temporal consistency. As shown in Tab.~\ref{table:comparison1}, our method achieves the highest rendering quality and surpasses existing methods in compression efficiency.

\begin{figure*}[t] 
	\begin{center} 
		\includegraphics[width=0.99\linewidth]{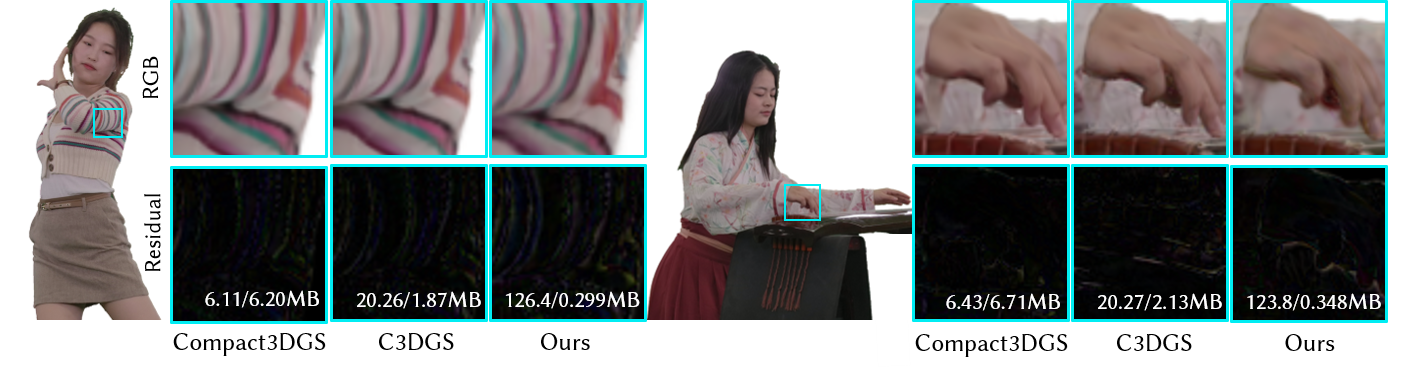} 
	\end{center} 
	\vspace{-16pt}
	\caption{\jyh{Qualitative comparison of our method against Compact-3DGS~\cite{Lee_2024_CVPR}, C3DGS~\cite{Niedermayr_2024_CVPR} on our challenging dataset. we calculate the residual map between the predictions and ground truth. Our method achieves the highest compression ratio while maintaining comparable rendering quality.}}
	\label{fig:fig_comp_2}
 	\vspace{-9pt}

\end{figure*} 

\begin{figure*}[t] 
	\begin{center} 
		\includegraphics[width=1\linewidth]{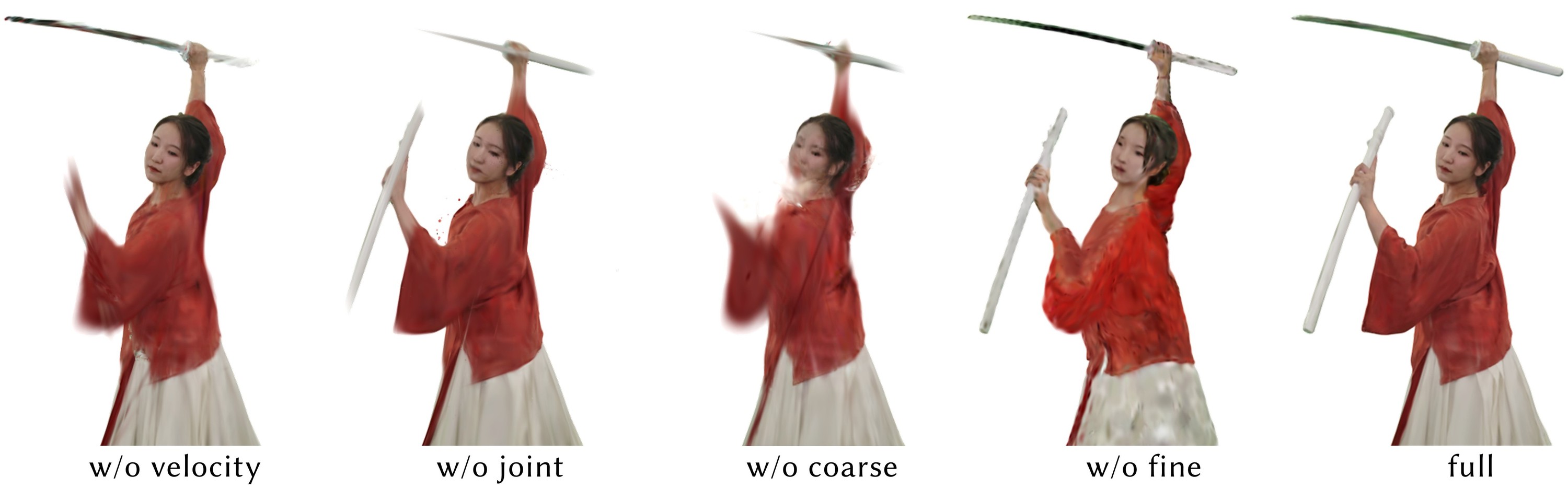}
	\end{center} 
	\vspace{-15pt}
	\caption{Qualitative evaluation of our Dual-Gaussian representation.} 
	\label{eval:dual gaussian}
	\vspace{-10pt}
\end{figure*}

\begin{figure*}[t] 
	\begin{center} 
		\includegraphics[width=0.99\linewidth]{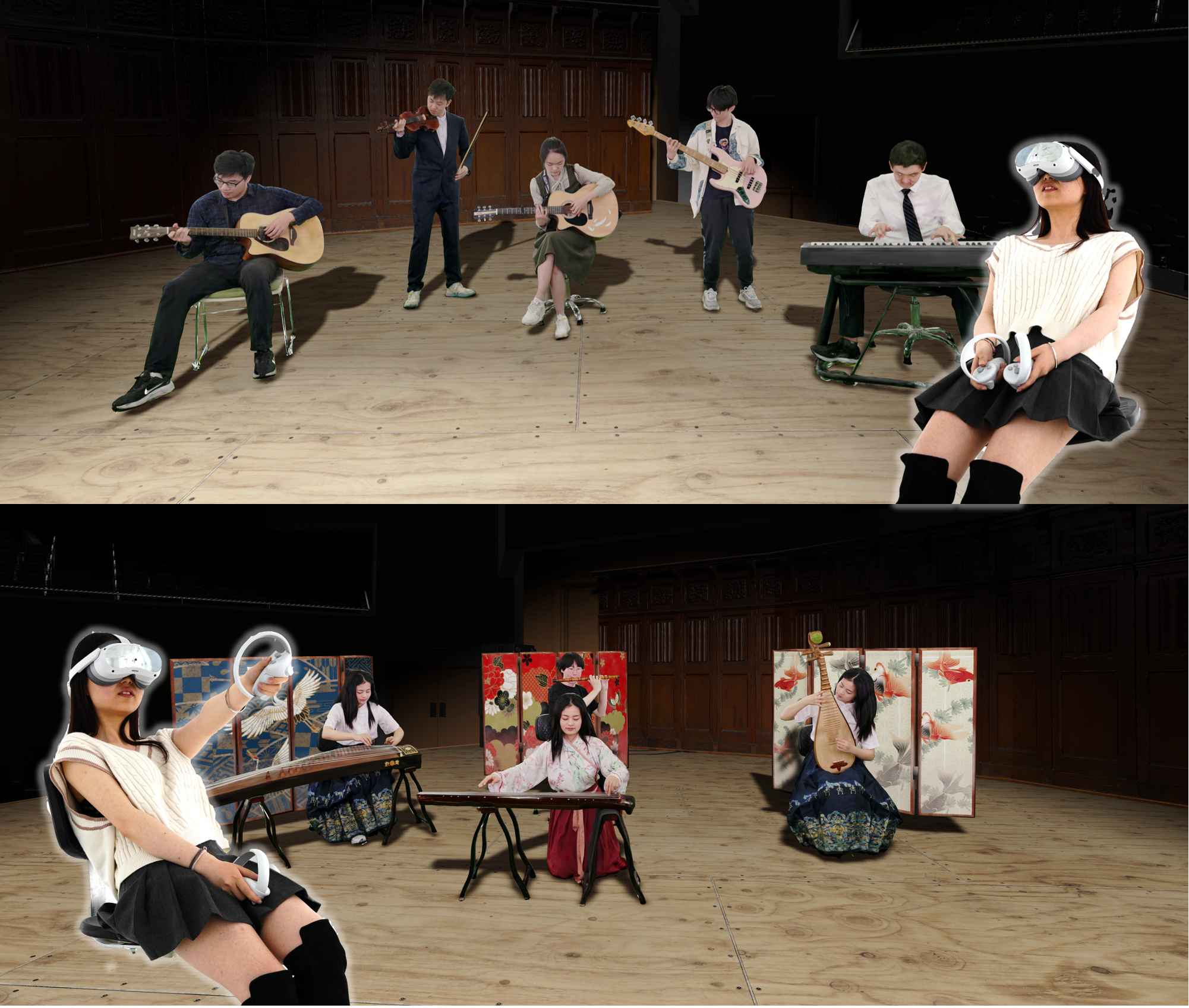} 
	\end{center} 
	\vspace{-5pt}
	\caption{We develop a VR demo to showcase how users can immerse themselves in a virtual musical odyssey, standing beside musicians, observing their performances up close, and feeling the rhythm of the music.}
	\label{fig:application_vr}
 	\vspace{-9pt}

\end{figure*} 

\jyh{
\paragraph{Compression Comparison.} We then compare our method with SOTA static compression methods, Compact-3DGS~\cite{Lee_2024_CVPR} and C3DGS~\cite{Niedermayr_2024_CVPR}. Since we apply these methods to our human performance dataset, the resulting compression ratios differ from those reported in their respective papers for static scenes. In addition to rendering RGB images, we also calculate the residual map between the predictions and ground truth.  As shown in Fig.~\ref{fig:fig_comp_2}, our DualGS compression achieves excellent storage efficiency while maintain the comparable rendering quality to other static Gaussian compression methods. 
A quantitative comparison is provided in Tab.~\ref{table:comparison2}, our method leverages spatial-temporal redundancy by using residual vector quantization to compress joint Gaussians motion, codec compression to encode the opacity and scaling of skin Gaussians, and a persistent codebook to represent spherical harmonics, achieving a compression ratio of up to 120 times and requiring only approximately 350KB of storage per frame.}

\begin{table}[t]
	\begin{center}
		\centering
		\vspace{-1pt}
        \caption{
        \jyh{
        \textbf{Quantitative comparison with static compression methods}. Green and yellow cell colors indicate the best and the second-best results.}}        
        \vspace{-2pt}
		\resizebox{0.47\textwidth}{!}{
			\begin{tabular}{l|cccccc}
				\hline
				Method   &  PSNR $\uparrow$ & SSIM $\uparrow$ & LPIPS $\downarrow$ & Running Time $\downarrow$  & Storage(MB / frame) $\downarrow$ \\
				\hline
			    Compact-3DGS~\cite{Lee_2024_CVPR}     & \colorbox{best1}{36.021} & \colorbox{best1}{0.991} & \colorbox{best2}{0.0193} & \colorbox{best2}{14m 21s} & 6.576 (6.38x) \\
				C3DGS~\cite{Niedermayr_2024_CVPR}  & \colorbox{best2}{35.823} & \colorbox{best2}{0.991} & \colorbox{best1}{0.0189} & 16m 09s & \colorbox{best2}{2.088 (20.11x)} \\
                Ours & 35.243 & 0.989 & 0.0221 & \colorbox{best1}{12m 12s} & \colorbox{best1}{0.323 (122.76x)}\\
				\hline
			\end{tabular}
		}
		\label{table:comparison2}
    
    	\vspace{-3pt}
	\end{center}
\end{table}

\begin{table}[t]
    \centering
    \caption{Compression Strategies on different attributes. Error is the mean absolute difference from the uncompressed data.}
    \resizebox{\columnwidth}{!}{ %
    \begin{tabular}{l|ccc|ccc}
    \hline
    \multirow{2}{*}{Method} & \multicolumn{3}{c|}{Total size (KB)} & \multicolumn{3}{c}{Error} \\ \cline{2-7}
     & Motion & OP+Scale & SH & Motion & OP+Scale & SH  \\ \hline
     Raw(PLY) & 462 & 2801 & 35573 & 0.0 & 0.0 & 0.0 \\
     Residual & \cellcolor{gray!50}96.3 & 362.3 & 1086.6 & 0.00124 & 0.20353 & 0.04265 \\
     Codec & 27.44 & \cellcolor{gray!50}126 & 226 & 0.01037 & 0.04661 & 0.00571 \\
     Codebook & 38.187 & 658.36 & \cellcolor{gray!50}98.76 & 0.03997 & 0.05519 & 0.01127 \\
    \hline
    \end{tabular}}
    \vspace{-12pt}
    \label{tab:compression}
\end{table}

\begin{table}[h]
\centering
\caption{Codebook sizes evaluation. Grey rows indicate our configurations.}
\label{tab:codebook_size}
\resizebox{\columnwidth}{!}{
\begin{tabular}{l|c|c|c|c}
\hline
\multirow{2}{*}{Codebook Size} & \multicolumn{4}{c}{Metrics} \\
\cline{2-5}
& PSNR $\uparrow$ & SSIM $\uparrow$ & LPIPS $\downarrow$ & SIZE (KB) $\downarrow$ \\
\hline
1024 & 35.447(-0.574) & 0.99372(-0.000948) & 0.02031(+0.00242) & 54.8 \\
2048 & 35.558(-0.463) & 0.99386(-0.000807) & 0.01999(+0.00210) & 63.4 \\
4096 & 35.653(-0.368) & 0.99396(-0.000708) & 0.01968(+0.00179) & 77.3 \\
\rowcolor{gray!50}
8192 & 35.729(-0.292) & 0.99404(-0.000628) & 0.01944(+0.00155) & 99.4 \\
16384 & 35.771(-0.250) & 0.99413(-0.000538) & 0.01914(+0.00125) & 152.09 \\
\hline

\end{tabular}
}
\vspace{-9pt}

\end{table}
\subsection{Ablations} \label{sec:abla} 

\paragraph{Dual Gaussian Representation.} We conduct a qualitative ablation study on the dual Gaussian representation to evaluate its efficacy. As shown in Fig.~\ref{eval:dual gaussian}, omitting the velocity prediction(w/o velocity) and relying solely on the smoothing term leads to inaccurate tracking during fast motions. Furthermore, the exclusion of the joint Gaussians(w/o joint) and optimizing all attributes of the skin Gaussians produces severe artifacts and loses temporal consistency. Additionally, omitting the coarse alignment stage(w/o coarse) introduces noticeable artifacts, and excluding the fine-grained optimization(w/o fine) yields unnatural outputs due to the fixed appearances, despite relatively accurate motion capture. In contrast, our full pipeline significantly enhances tracking accuracy, ensuring temporal consistency and achieving high-fidelity rendering.

\paragraph{Hybrid Compression Strategy.}
As shown in Tab.~\ref{tab:compression}, 
we evaluate three compression strategies for different attributes. 
For the motion attributes (position and rotation) of joint Gaussians, residual compression provides the highest precision with acceptable size. For the opacity and scaling of skin Gaussians, codec compression achieves optimal precision with minimal storage requirements. For the storage-intensive SH attribute, we balance precision with storage efficiency by using a persistent codebook for compression.

\paragraph{Codebook Size.}
Spatial-temporal SH coefficients are compressed into persistent codebooks of predefined sizes. As shown in Tab.~\ref{tab:codebook_size}, enlarging the codebook size beyond 8,192 yields little effect in compression efficiency, while significantly increasing storage consumption. Consequently, we keep the storage overhead of SH coefficients at levels comparable to those required for storing XYZ coordinates.

\jyh{
\subsection{Immersive Experiences} \label{sec:experiences}
In Fig.~\ref{fig:application_vr}, we showcase the application of watching a high-fidelity virtual concert using the PICO 4 VR headset. Viewers can immerse themselves in a virtual musical experience, observing musicians perform up close as if standing beside them, even though the musicians are located in various places around the world. We can capture, process, and generate 4D assets at different times, integrating individual performers into a consistent environment within a standard CG engine. Additionally, we can edit their positions and align their actions temporally, ensuring that their performances are visually synchronized with the rhythm of the ensemble.}

\jyh{
\subsection{Limitations and Discussions} \label{sec:limit}
Our approach achieves template-free dynamic human modeling via the disentangled and hierarchical motion and appearance representation. With the specially designed compression strategy, we achieve a 120-fold compression ratio and still deliver accurate tracking and high-fidelity rendering for immersive experiences. Despite such compelling capabilities, our pipeline still yields some limitations. We provide detailed analysis and discuss potential future extensions.

Firstly, our method relies on image-based accurate segmentation to separate the foreground human performances, which may yield segmentation errors with slender objects such as lute strings and hair, compromising detailed tracking. It is an interesting direction to incorporate the view-consistent matting and 3D/4D understanding to enhance 4D modeling and rendering. 
Moreover, although our DualGS representation avoids explicit mesh reconstruction and non-rigid tracking, it still requires more training time compared to real-time tracking. The bottleneck lies in the coarse alignment and fine-grained optimization. To further accelerate the process, we observe that using 180,000 skin Gaussians to represent human appearance may introduce redundancy. A potential solution is to employ LightGaussian~\cite{fan2023lightgaussian} to prune unimportant Gaussian kernels, thereby speeding up the rendering process as well.
To ensure efficient motion tracking and temporal consistency, we fix the joint-skin KNN relationship after initialization. However, this sacrifices the ability to handle topological changes. Combing dynamic graph~\cite{FlyFusion} and keyframe strategy~\cite{dou2016fusion4d} may address this issue.
Our method can produce vivid volumetric videos without relying on human parametric models or skeleton information. However, it does not support downstream tasks such as animatable avatar or motion transfer. Annotating 4D sequences and driving the 4D assets using multimodal inputs, such as text prompts, music, or human skeleton, is promising. Moreover, although we already integrate our 4D assets into standard CG engines, the lack of geometric or normal information prevents them from being re-lit under different lighting conditions, which presents an interesting avenue for future research.}

\section{CONCLUSION}

We have presented a comprehensive solution for producing high-fidelity, human-centric volumetric video. Our core approach is based on a dual-Gaussian representation for challenging human performance, enabling accurate tracking and high-fidelity rendering. By organically combining a compact number of motion-aware joint Gaussians to capture global movements with a larger set of appearance-aware skin Gaussians for visual details, we adeptly manage challenging motions without sacrificing quality. 
\jyh{For Dual-Gaussian initialization, we utilize a uniform random point cloud to initialize the joint Gaussians and carefully control their number and scale. These joint Gaussians serve as the foundation for initializing the skin Gaussians and constructing the KNN field for subsequent optimization.} Furthermore, we propose a coarse-to-fine training strategy to reduce optimization difficulty. To integrate long volumetric video sequences into VR platforms, we have developed a DualGS-based compression strategy to achieve a 120-fold compression ratio. \jyh{We also implement a companion Unity plugin for hybrid rendering with a standard CG immersive environment as well as DualGS player that enables high-quality rendering on low-end mobile devices.}
Experimental results demonstrate that our method vividly produces high-quality renderings. We believe our method serves as a "ticket" to a virtual world, offering immersive and high-fidelity experiences.

\noindent{\bf Acknowledgements.} We thank reviewers for their feedback. This work was supported by  NSFC programs (61976138), National Key R$\&$D Program of China (2022YFF0902301), STCSM (2015F0203-000-06), Shanghai Local college capacity building program (22010502800). We also acknowledge support from Shanghai Frontiers Science Center of Human-centered Artificial Intelligence (ShangHAI).

\bibliographystyle{ACM-Reference-Format}
\bibliography{sample-bibliography}

\end{document}